\documentstyle[sprocl]{article}
\input{psfig}
\def\be{\begin{equation}}
\def\ee{\end{equation}}
\def\bea{\begin{eqnarray}}
\def\eea{\end{eqnarray}}

\def\approxlt{\ifmmode \rlap{$<$}{}_{{}_{{}_{\textstyle\sim}}} \else%
$\rlap{$<$}{}_{{}_{{}_{\textstyle\sim}}}$\fi}
\def\msun{\ifmmode \rm{M}_\odot \else M$_\odot$\fi}	% solar mass
\def\mdot{\ifmmode \dot M \else $\dot M$\fi}	% accretion rate
\def\mc{\multicolumn}
\def\da{$\downarrow$}
\begin{document}

\title{MILLISECOND PHENOMENA IN ACCRETING NEUTRON STARS -- AN UPDATE}
\author{M. VAN DER KLIS}
\address{Astronomical Institute ``Anton Pannekoek'', University of
Amsterdam\\Kruislaan 403, 1098 SJ Amsterdam, The Netherlands\\E-mail: 
michiel@astro.uva.nl}
%\author{ A.N. OTHER }
%\address{Department of Physics, Theoretical Physics, 1 Keble Road,\\
%Oxford OX1 3NP, England\\E-mail: other@tp.ox.uk}
% You may repeat \author \address as often as necessary      

\maketitle

\abstracts{Since the initial discoveries with the Rossi X-ray Timing
Explorer (RXTE) in 1996 of kilohertz quasi-periodic oscillations (kHz
QPOs) and burst oscillations in a number of low-mass X-ray binaries
(LMXBs) containing low-magnetic-field neutron stars, a very active
field has developed. I briefly summarize some of the developments
since those early days, which include the discovery of the first
accreting millisecond pulsar, further claims of the detection of
strong-field general-relativistic effects, intriguing correlations of
the kHz QPO properties with QPOs at lower frequencies, strong
challenges for the beat-frequency interpretation of the twin kHz QPO
peaks and a number of new theoretical ideas, some of which involving
frame dragging. Twenty LMXBs have now been seen to exhibit periodic or
quasi-periodic phenomena with frequencies exceeding
10$^{2.5}$\,Hz. Although the commensurabilities between the observed
frequencies that suggest a beat-frequency interpretation have
conclusively been shown to be not precise, the preponderance of the
evidence still is in favour of the idea that there exists some kind of
beat-frequency relation between the twin kHz peaks. However, that
evidence is only coming from 4 of the 17 sources showing twin kHz
QPOs.}

\section{Introduction}\label{sect:intro}

In this review, I attempt to summarize what happened in the field of
millisecond X-ray variability from accreting low-magnetic field neutron
stars in the period of slightly over one year since I submitted my
review for the proceedings of the Lipari NATO ASI (van der Klis
1998). So, for anything that happened before October 1997 and for much
of the introductory material, I refer to that article. As in that
article, I refer to the tables summarizing the phenomenology for most
of the references. A brief account of what went before:

Twin kilohertz quasi-periodic oscillation (QPO) peaks began to be
discovered with the Rossi X-ray Timing Explorer (RXTE) early
1996, briefly after its launch. The peaks had associated rms
amplitudes of $\approxlt$1\% up to several 10\% of the total source
flux, peak separations of $\sim$250--360\,Hz, coherences between Q
$\sim$ 1--10$^2$, and their frequencies (of the higher-frequency peak of the
two) varied between 500 and 1200\,Hz with mass accretion rate
\mdot. Only a short time later another series of RXTE discoveries
began, of a different type of rapid oscillations. These came from some
of the same sources, but only during type 1 X-ray bursts, and had
frequencies between 330 and 590 Hz, depending on the source. These
burst oscillations were interpreted in terms of the spin frequency of
a layer in the neutron star atmosphere that is presumably close to the
neutron star spin frequency itself. Finally, quite recently, the first
true spin frequency was detected (at 401 Hz) in an accreting
low-magnetic field neutron star.  In several sources it was found that
the twin kHz peak separation was equal to the frequency of the burst
oscillations, or half that, to within a few percent. This strongly
suggested a beat-frequency interpretation, where the higher-frequency
peak (hereafter the upper peak) is identified with orbital motion at
some preferred radius in the disk and the lower-frequency kHz QPO peak
(hereafter the lower peak) is at the beat frequency between this
orbital frequency and the spin frequency of the neutron star:
$\nu_{lo} = \nu_{upp} - \nu_s$. It was immediately realized that if we
are indeed seeing the signature of orbital motion around neutron stars
with orbital frequencies of up to 1200\,Hz, then we are probing a
region of space-time where strong-field general relativity is required
to describe orbital motion, and constraining the mass-radius relation
of neutron stars and thereby the equation of state (EOS) of
supranuclear-density matter.  This spurred a large amount of both
observational and theoretical work on these phenomena, some of which I
will summarize here.

A few ballpark numbers are useful to keep in mind. Orbital motion
around a $M$=1.4\msun\ neutron star with a frequency $\nu_K$ of
1200\,Hz takes place at an orbital radius of $r_K$=15\,km; the
expressions are $$\nu_K =\left(GM\over4\pi^2r_K^3\right)^{1/2} \approx
1200\,\hbox{Hz}\,\left(r_K\over15\,\hbox{km}\right)^{-3/2}\left(M\over1.4\msun\right)^{1/2},$$
or $$r_K = \left(GM\over4\pi^2\nu_K^2\right)^{1/3} \approx
15\,\hbox{km}\,\left(\nu_K\over1200\,\hbox{Hz}\right)^{-2/3}\left(M\over1.4\msun\right)^{1/3}$$
(this is exact for a Schwarzschild geometry and as measured at
infinity; for realistic neutron stars the frame dragging corrections
are moderate). The radius of the innermost stable circular orbit
(ISCO) calculated in the same geometry is $R_{ISCO}=6GM/c^2$, which
for 1.4\msun\ works out to 12.5\,km.

\section{Phenomenology}

\subsection{Millisecond pulsar}\label{subsect:mspulsar}

\begin{figure}[ht]
\begin{center}
$$\psfig{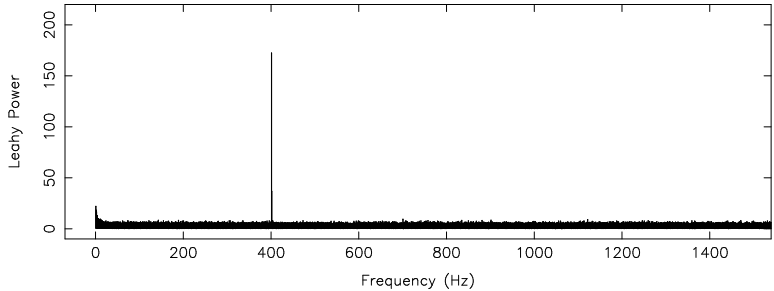}$$
\end{center}
\caption{The discovery power spectrum of the first accreting
millisecond X-ray pulsar. (Wijnands and van der Klis 1998b)\label{fig:1808fft}}
\end{figure}

The first accreting millisecond pulsar was finally discovered, more
than two years after RXTE's launch, by Rudy Wijnands using RXTE on
April 13, 1998 in the soft X-ray transient SAX\,J1808.4$-$3658
(Fig.~\ref{fig:1808fft}; Wijnands and van der Klis 1998a,b). The pulse
frequency is 401\,Hz.

This discovery, a long sought after ``Holy Grail'' of X-ray astronomy,
had been anticipated for nearly 20 years, because magnetospheric disk
accretion theory as well as the example of the non-accreting
millisecond {\it radio} pulsars indicated that such rapid spin
frequencies had to exist among accreting low-magnetic field neutron
stars -- yet despite numerous searches such rapid pulsars had not
turned up in observations of LMXBs. Now that one such accreting
millisecond pulsar is known, of course, the question becomes ``why
only one?''. The pulsations in SAX\,J1808.4$-$3658 are not
particularly weak and we know for sure that pulsars of similar
amplitude and observed (as opposed to intrinsic) coherence are not
present in many other LMXBs observed in a similar (rather standard)
way as SAX\,J1808.4$-$3658 was with RXTE.  The answer to this question
may lie in part in the orbital characteristics of the source. The
2.0-hr orbital period was discovered, and the orbit measured, with
admirable speed by Chakrabarty and Morgan (1998a,b). With a projected
orbital radius $a\sin i$ of only 63 light {\it milli}seconds and a
mass function of 3.8\,10$^{-5}$\msun, the companion is either very low
mass, or we are seeing the orbit nearly pole-on.

\begin{figure}[ht]
$$\psfig{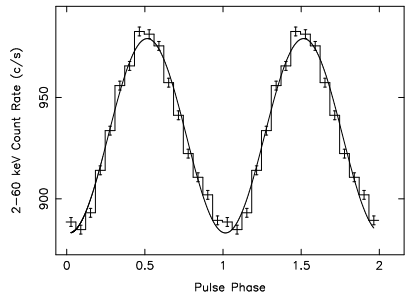}$$
\caption{The pulse profile of SAX\,J1808.4$-$3658. (Wijnands and van
der Klis 1998b)\label{fig:1808profile}}
\end{figure}

If we see the orbit pole-on, then this may be what allows us to see
the pulsations in this system, and a different orientation of the
orbit what hides them in many other LMXBs. The fact that the pulse
profile of SAX\,J1808.4$-$3658 (Fig.~\ref{fig:1808profile}) is nearly
sinusoidal might be related to this pole-on orientation. In any case,
of course, the low radial-velocity amplitude of the orbit reduces the
Doppler shifts and keeps the pulsar more nearly coherent, facilitating
its discovery without the use of special deacceleration techniques
(which are, however, being used to try and discover other pulsars;
e.g., Vaughan et al. 1994; van der Klis et al. 1999, in prep.).

Another possibility is that SAX\,J1808.4$-$3658 is a different type of
system from the other well-studied LMXBs. Perhaps the fact that it is
an unspectacular little transient is related to an accretion history
that favours the preservation of a magnetic field configuration
(strength, orientation) that produces a pulsar, and some {\it other}
unspectacular little transients (with, perhaps, low-mass companions)
will turn out to be millisecond pulsars as well.

It would be of enormous interest to find burst oscillations
(\S\ref{subsect:burstosc}) or kHz QPOs (\S\ref{subsect:khzqpo}) in
SAX\,J1808.4$-$3658, as this would allow right away to confirm or
reject various models for these phenomena involving the neutron star
spin, which in SAX\,J1808.4$-$3658, uniquely among accreting
low-magnetic field neutron stars, is known accurately and with
certainty. However, the RXTE observations during the first few days of
the April 1998 outburst, when the source was brightest, were
relatively limited. Perhaps for this reason, although
SAX\,J1808.4$-$3658 is known from its September 1996 SAX discovery
observation (in 't Zand et al. 1998) to be a type 1 X-ray burster, no
bursts were seen in the RXTE observations of the April 1998 outburst
(the source was unobservable for SAX during that time), nor were kHz
QPOs detected.  Early in the outburst, when from comparing to other
LMXBs the chances of seeing kHz QPOs were best, due to the limited
observing time RXTE was not very sensitive to them (Wijnands and van
der Klis 1998c). Plans are in place to spend a large amount of
observing time with RXTE on SAX\,J1808.4$-$3658 during its next
outburst, which should allow us to do a good job with respect to this
and various other issues related to this interesting source.

\subsection{Burst oscillations}\label{subsect:burstosc}

Six sources have now shown burst oscillations.  The frequencies are
listed in Table~\ref{tab:burstosc}. These oscillations are not
detected in each burst.  In four of these sources (the top four in
Table~\ref{tab:burstosc}), twin kHz peaks in the persistent emission
have also been observed (see \S\ref{subsect:khzqpo}). It turns out
that in each case, the burst QPO frequencies are close to the
frequency {\it differences} between the twin peaks, or twice
that. These frequencies have been listed in Table~\ref{tab:burstosc}
in the columns labeled Frequency 1 and Frequency 2, respectively.

\begin{table}[hb]
\footnotesize
\caption{Burst oscillations\label{tab:burstosc}} 
\begin{center}
\footnotesize
\begin{tabular}{|lcc|}
\hline 
Source        & Frequency 1 & Frequency 2 \\
              & (Hz)        & (Hz)        \\
\hline
4U\,1636$-$53 & 290         & 581         \\
4U\,1702$-$43 & 330         & ---         \\
4U\,1728$-$34 & 363         & ---         \\
KS\,1731$-$260& ---         & 524         \\
MXB\,1743$-$29& \mc{2}{c|}{589}            \\
Aql\,X-1      & \mc{2}{c|}{549}            \\
\hline
\end{tabular}
\end{center}
References see Tables~\ref{tab:z}, \ref{tab:atola} and
\ref{tab:atolb}. Source identification of MXB\,1743$-$29 is
uncertain. Frequency 1 is close to the kHz QPO peak separation
frequency, frequency 2 to twice that. For MXB\,1743$-$29 and Aql\,X-1
this separation frequency is unknown.
\end{table}

\begin{figure}[ht]
$$\psfig{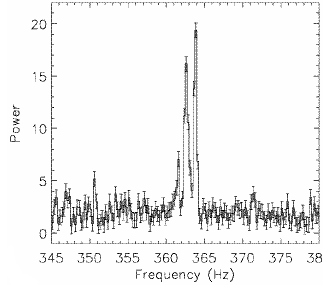}$$
\caption{A drifting burst oscillation in 4U\,1728$-$34. (Stromayer et
al. 1996c)\label{fig:burstqpo}}
\end{figure}

The burst QPOs have a relatively large coherence. In KS\,1731$-$260 a Q
of 900 was reported (Smith, Morgan and Bradt 1997). However, usually
drifts by a few Hz are observed in the QPO frequency (Strohmayer et
al. 1998c; Fig.~\ref{fig:burstqpo}). These drifts are suggestive of
the bursting layer slightly expanding and then recontracting, changing
its rotation rate to conserve angular momentum and thus modulating the
QPO frequency. In 4U\,1728$-$34 the QPO amplitude decreases and the
fitted black-body radius increases during the burst rise in a way that
is very suggestive of a modulation caused by an expanding hot spot
(Strohmayer, Zhang and Swank 1997b). The high Q value, the small
frequency drifts and the amplitude variations of the burst QPOs
therefore support the view that their origin lies in the neutron star
spin, and that their frequency is close to the spin frequency.

\begin{figure}[ht]
$$\psfig{figure=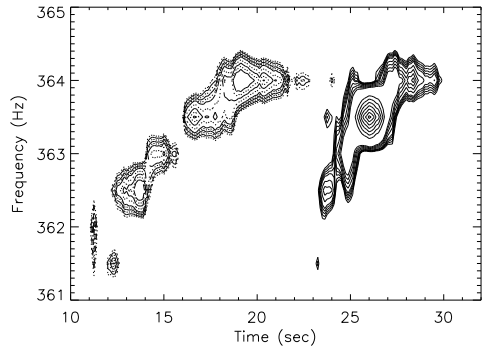,height=2in}$$
\caption{Dynamic power spectrum of burst
oscillations in two bursts separated by 1.6\,yr in 4U\,1728$-$34 showing near-identical asymptotic
frequencies. (Strohmayer et al. 1998b) \label{fig:burstasymp}}
\end{figure}

In addition to this evidence, it has been shown by Strohmayer et
al. (1998b) that the frequency to which the oscillation approaches
near the end of the burst (Fig.~\ref{fig:burstasymp}) is constant to
within about 0.01\% between bursts far apart in time. In the above
interpretation this asymptotic frequency is the rotation frequency of
the bursting layer when it approaches its quiescent rotation state,
one that may be more nearly in corotation with the underlying
star. This is another argument in favour of the near-constant neutron
star spin as the process underlying the observed, slightly drifting
frequencies.

Miller (1998), from an analysis of five bursts in 4U\,1636$-$53,
reports the presence of a 290\,Hz burst oscillation at half the
frequency of, and simultaneous with, the stronger 580\,Hz oscillation,
and interprets this in terms of the presence of two antipodal hot
spots on the surface. If correct, this would indicate a neutron star
spin frequency of 290\,Hz (more nearly equal to the kHz peak
separation, see \S\ref{subsect:khzqpo}), and strongly suggests we are
seeing the signature of the presence of two magnetic poles. As the
oscillation frequency drifts, we can {\it not} be looking at the
magnetic poles themselves, which are of course fixed in the frame
corotating with the spin of the star. Perhaps we are seeing the effect
of the accumulation of extra nuclear fuel at the poles, which becomes
decoupled from the magnetic field lines by convection during the
bursts (Bildsten 1996, priv. comm.). The required near-simultaneous
ignition of both poles at the burst onset may be a problem for this
interpretation.

\subsection{Kilohertz quasi-periodic oscillations} \label{subsect:khzqpo}

\begin{figure}[ht]
$$\psfig{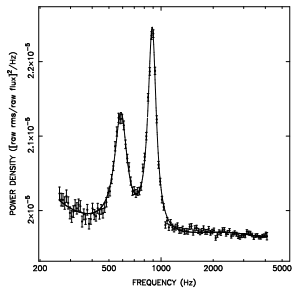}$$
\caption{Twin kHz peaks in Sco\,X-1. (van der Klis et al. 1997b) \label{fig:twinpeaks}}
\end{figure}

The number of sources that have shown kilohertz quasi-periodic
oscillations (kHz QPOs) is now\footnote{December 21, 1998}
eightteen. Together with the one source that showed burst oscillations
(\S\ref{subsect:burstosc}) but no kHz QPOs, and the millisecond pulsar
(\S\ref{subsect:mspulsar}), which so far showed neither, this brings
the total number of sources that has shown kilohertz periodic or
quasi-periodic phenomena (here defined as phenomena at frequencies
exceeding 10$^{2.5}$\,Hz) to twenty. All but one of the kHz QPO
sources have shown twin kHz peaks in their power spectrum
(Fig.~\ref{fig:twinpeaks}); the exception with only a single peak so
far is Aql\,X-1.  Tables~\ref{tab:z}, \ref{tab:atola} and
\ref{tab:atolb} summarize the results on these sources.

\begin{table}[htbp]
\caption{Observed frequencies of kilohertz QPOs in Z sources} 
\label{tab:z}
\footnotesize
\begin{tabular}{|lccccl|}
\hline
Source & Lower & Upper & Peak & Burst & References\\ & peak & peak &
                sepa- & osc.&\\ & freq.  & freq.  & ration & freq.&\\
                & (Hz) & (Hz) & (Hz) & (Hz)&\\
\hline 
Sco\,X-1        & 565    & 870     & 307$\pm$5 && Van der Klis et al. 1996a,b,c,1997b \\
                & \da    & \da     & \da       && \\
                & 845    & 1080    & 237$\pm$5 && \\
                &        & \da     &           && \\
                &        & 1130    &           && \\
\hline					       
GX\,5$-$1       & 215    & 505     &           && Van der Klis et al. 1996e \\
                & \da    & \da     & 298$\pm$11&& Wijnands et al. 1998c \\
                & 660    & 890     &           && \\
                & \da    &         &           && \\
                & 700    &         &           && \\
\hline					       
GX\,17+2        &        & 645     &           && Van der Klis et al. 1997a \\
                &        & \da     &           && Wijnands et al. 1997b \\
                & 480    & 783     &           && \\
                & \da    & \da     & 294$\pm$8 && \\
                & 780    & 1080    &           && \\
                &        & \da     &           && \\
                &        & 1085    &           && \\
\hline
Cyg\,X-2        &        & 730     &           && Wijnands et al. 1998a \\
                &        & \da     &           && \\
                & 530    & 855     &           && \\
                & \da    & \da     & 346$\pm$29&& \\
                & 660    & 1005    &           && \\
\hline
GX\,340+0       & 250    & 570     &           && Jonker et al. 1998 \\
                & \da    & \da     & 325$\pm$10&& \\
                & 500    & 820     &           && \\
                & \da    &         &           && \\
                & 625    &         &           && \\
\hline
GX\,349+2       & 710    & 980     & 266$\pm$13&& Zhang et al. 1998a \\
                &\mc{2}{c}{1020$^a$}&          && Kuulkers and van der Klis 1997 \\
\hline
\end{tabular}
\vbox{These notes refer also to Tables~\ref{tab:atola} and \ref{tab:atolb}.
Frequencies with no quoted errors were rounded to the nearest 5, or,
for burst QPOs, 1\,Hz.  Arrows indicate ranges over which the frequency
was observed or inferred to vary; these can be made up of several
overlapping ranges from different observations.  Frequencies not
connected by arrows are measurements at different epochs.  Frequencies
in the same row were observed simultaneously (except for burst QPOs).
Entries straddling the upper and lower peak columns are of single,
unidentified peaks.  $^a$ Marginal detection.  $^b$ Special detection
method; M\'endez et al. (1998a,b,c).  $^c$ Source identification
uncertain.  $^d$ Pulsar.}
\end{table}

\begin{table}[htbp]
\caption{Observed frequencies of kilohertz QPOs in atoll sources - Part I} 
\label{tab:atola}
\footnotesize
\begin{tabular}{|lccccl|}
\hline
Source          & Lower  & Upper   & Peak       & Burst       & References\\
                & peak   & peak    & sepa-      & osc.&\\
                & freq.  & freq.   & ration     & freq.&\\
                & (Hz)   & (Hz)    & (Hz)       & (Hz)&\\ 
\hline					       
4U\,0614+09     &        & 540     &            &             & Ford et al. 1996, 1997a,b \\
                &        & \da     &            &             & Van der Klis et al. 1996d \\
                & 485    & 840     &            &             & M\'endez et al. 1997 \\
                & \da    & \da     & 323$\pm$4  &             & Vaughan et al. 1997,1998 \\
                & 800    & 1145    &            &             & Kaaret et al. 1998 \\
\hline
4U\,1608$-$52   & \mc{2}{c}{270}   &            &             & Van Paradijs et al. 1996 \\
                & 440    & 765     & 325$\pm$7  &             & Berger et al. 1996 \\
                & \da    & \da     & \da        &             & Vaughan et al. 1997, 1998 \\
                & 475    & 800$^b$ & 326$\pm$3  &             & Yu et al. 1997 \\
                & \da    & \da     & \da        &             & M\'endez et al. 1998a,b,d \\
                & 865    & 1090$^b$& 225$\pm$12 &             & Kaaret et al. 1998 \\
                & \da    &         &            &             & \\ 
                & 895    &         &            &             & \\
\hline
4U\,1636$-$53   & 830    &         &            &             & Zhang et al.  1996, 1997a \\ 
                & \da    &         &            &             & Zhang 1997 \\
                & 900    & 1150    & 257$\pm$20 &             & Van der Klis et al. 1996d \\ 
                & \da    & \da     &            &             & Wijnands et al. 1997a \\ 
                & 950    & 1190    & 276$\pm$10 &             & Vaughan et al. 1997 \\ 
                &        & \da     &            &             & Miller 1998 \\
                &        & 1228    &            &             & Strohmayer et al. 1998b \\
                &        &         &            &             & \\
                & 830    &         &            &             & M\'endez et al. 1998c \\
                & \da    & 1190$^b$& 251$\pm$4$^b$& 290,581   & \\
                & 1050   &         &            &             & \\
\hline
4U\,1702$-$43   & 625    &         &            &             & Markwardt et al. 1998 \\
                & \da    &1058$^b$ &333$\pm$5$^b$& 330         & \\
                & 825    &         &            &             & \\
                &        &         &            &             &\\
                & 655    & 1000$^b$&344$\pm$7$^b$&             & \\
                & \da    & \da     & \da        &             & \\
                & 700    & 1040$^b$&337$\pm$7$^b$&             & \\
                & \da    & \da     & \da        &             & \\
                & 770    & 1085$^b$&315$\pm$11$^b$&             & \\
                & 902    &         &            &             & \\
\hline
4U\,1705$-$44   & 775    & 1075$^a$& 298$\pm$11 &             & Ford et al. 1998a \\
                & \da    &         &            &             & \\
                & 870    &         &            &             & \\
\hline
\end{tabular}
Notes: see Table~\ref{tab:z}.
\end{table}

\begin{table}[htbp]
\caption{Observed frequencies of kilohertz QPOs in atoll sources - Part II} 
\label{tab:atolb}
\footnotesize
\begin{tabular}{|lccccl|}
\hline
Source          & Lower  & Upper   & Peak       & Burst       & References\\
                & peak   & peak    & sepa-      & osc.&\\
                & freq.  & freq.   & ration     & freq.&\\
                & (Hz)   & (Hz)    & (Hz)       & (Hz)&\\ 
\hline
4U\,1728$-$34   &        &  325$^f$&            &             & Strohmayer et al. 1996a,b,c,\\
                &        & \da     &            &             & 1997b,1998b \\
                &        &  500    &            &             & Ford and van der Klis 1998 \\
                &        & \da     &            &             & \\
                & 640    &  990    &            &             & \\
                & \da    & \da     & 355$\pm$5  & 364         & \\
                & 790    & 1120    &            &             & \\
\hline					       
KS\,1731$-$260  & 900    & 1160    & 260$\pm$10 & 524         & Morgan and Smith 1996 \\
                &        & \da     &            &             & Smith et al. 1997, \\
                &        & 1205    &            &             & Wijnands and Van der Klis 1997 \\
\hline
4U\,1735$-$44   & 640    & 980     & 341$\pm$7  &             & Wijnands et al. 1996, 1998b \\
                & \da    & \da     & \da        &             & Ford et al. 1998b \\
                & 730    & 1025    & 296$\pm$12 &             & \\
                &        &         &            &             & \\
                & 900$^a$& 1150    &  249$\pm$15&             & \\ 
                &        & \da     &            &             & \\
                &        & 1160    &            &             & \\
\hline
MXB\,1743$-$29$^c$&      &         &            & 589         & Strohmayer et al. \\
                &        &         &            &             & 1996d,1997a \\
\hline
SAX\,J1808.4-3658&       &         &            & 401$^d$     & Wijnands
and van der Klis \\
                 &       &         &            &             & 1998a,b,c \\
\hline
4U\,1820$-$30   &        & 660     &            &             & Smale et al. 1996, 1997 \\
                &        & \da     &            &             & Zhang et al. 1998b \\
                & 500    & 860     & 358$\pm$43 &             & \\
                & \da    & \da     & \da        &             & \\
                & 796    & 1065    & 275$\pm$8  &             & \\
\hline
Aql\,X-1        &\mc{2}{c}{\ 677}  &            &             & Zhang et al. 1998c \\
                &\mc{2}{c}{\da}    &            & 549         & Cui et al. 1998  \\
                &\mc{2}{c}{\ 871}  &            &             & Yu et al. 1998  \\
\hline
4U\,1915-05     &        & 820     &            &             & Barret et al. 1997,1998 \\
                &        & \da     &            &             & \\
                & 555    & 935     &            &             & \\
                & \da    & \da     & 355$\pm$?  &             & \\
                & 655    & 1005    &            &             & \\
\hline
XTE\,J2123$-$058& 850    & 1110$^a$    & 261$\pm$10$^a$&         & Homan et al. 1998a,b \\
                & \da    & \da         & \da        &         & \\
                & 855    & 1123        & 268$\pm$9  &             & \\
                & \da    & \da         & \da        &         & \\
                & 870$^a$& 1140        & 270$\pm$5$^a$&         & \\
\hline
\end{tabular}
Notes: see Table~\ref{tab:z}.
\end{table}

\begin{figure}[ht]
$$\psfig{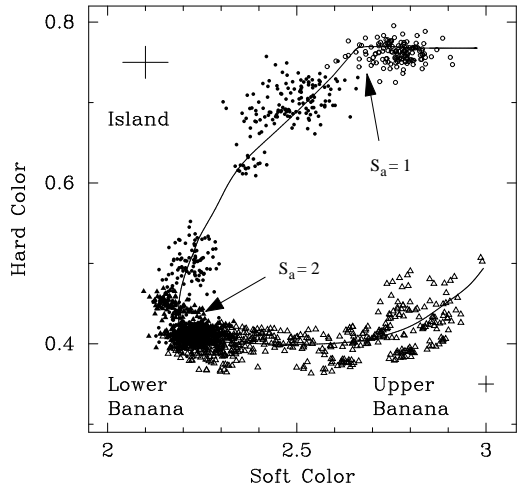}$$
\caption{X-ray color-color diagram of 4U\,1608$-$52. Mass accretion
rate is inferred to increase counterclockwise, approximately along the
drawn curve. kHz QPO detections are indicated with filled
symbols. (M\'endez et al. 1998d)\label{fig:1608cd}}
\end{figure}

The two kHz QPO peaks increase in frequency with inferred mass
accretion rate \mdot\ both in Z and in atoll sources (see Hasinger and
van der Klis 1989 or the review by van der Klis 1995 for the
introduction of these subtypes of LMXBs). In Z sources, the QPOs are
nearly always seen down to the lowest inferred \mdot\ levels these
sources reach, in atoll sources the QPOs tend to occur in the middle
of the \mdot\ range of each source (Fig.~\ref{fig:1608cd}). Kilohertz
QPOs are seen in a similar frequency range (500--1200\,Hz for the
upper peak) in sources that differ in average X-ray luminosity $L_x$
by 2.5 orders of magnitude (where $L_x$ is here defined simply as
$4\pi d^2f_x$ with $f_x$ the X-ray flux and $d$ the distance), and the
kHz QPO frequency seems to be determined more by the difference
between average and instantaneous $L_x$ of a source than by $L_x$
itself. This is unexplained and must mean that another parameter than
\mdot, related in some way to the average $L_x$, affects the QPO
frequency (van der Klis 1997). Perhaps this parameter is the neutron
star magnetic field strength.

In 4 sources, both twin kHz peaks and burst oscillations are seen, and
the fact that the burst oscillation frequency is near 1 or 2 times the
kHz peak separation frequency is the main argument for the beat
frequency interpretation of kHz QPOs. The evidence for this is
summarized in Table~\ref{tab:beatevidence}. In 4U\,1636$-$53,
M\'endez et al. (1998c) have shown that the correspondence is not
exact. Yet, although we would dearly like to have a few more examples,
the preponderance of the evidence still seems to indicate that at
least an approximate beat-frequency relation exists between the three
frequencies in those sources where they are all observed.

\begin{table}[htb]
\footnotesize
\caption{Commensurability of kHz QPOs and burst oscillation frequencies\label{tab:beatevidence}} 
\begin{center}
\footnotesize
\begin{tabular}{|lccc|}
\hline 
Source        & Burst oscillation & kHz QPO separation & Ratio \\
              & frequency         & frequency          & (burst/separation\\ 
              & (Hz)              & (Hz)               & frequency)\\
\hline
4U\,1636$-$53 & 290               & 251$\pm$4          & 1.155$\pm$0.018 \\
~~~~~~``      & 581               & ~``~~~~``          & 2.315$\pm$0.037 \\
4U\,1702$-$43 & 330               & 333$\pm$5          & 0.991$\pm$0.015\\
4U\,1728$-$34 & 364               & 355$\pm$5          & 1.025$\pm$0.014\\
KS\,1731$-$260& 524               & 260$\pm$10         & 2.015$\pm$0.078\\
\hline
\end{tabular}
\end{center}
References see Tables~\ref{tab:atola} and \ref{tab:atolb}.
\end{table}

\begin{figure}[ht]
$$\psfig{figure=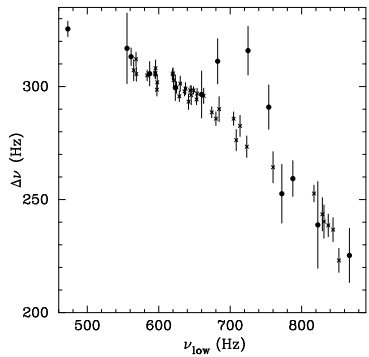,height=2in,angle=-90}$$
\caption{The variations in kHz QPO peak separation in Sco X-1 and
4U\,1608$-$52 as a function of the lower peak frequency. (M\'endez et
al. 1998b)\label{fig:sco1608}}
\end{figure}

In an interpretation where the burst oscillations and the kHz peak
separation are both close to the neutron star spin, this peak separation
is predicted to be approximately constant. It is clear that this is not
always the case: both in Sco\,X-1 and in 4U\,1608$-$52 this separation
decreases systematically when the kHz QPO frequencies increase
(Fig.~\ref{fig:sco1608}), and indications for this exist in
4U\,1735$-$44 and 4U\,1702$-$43 as well (see Tables~\ref{tab:atola} and
\ref{tab:atolb}).

\begin{figure}[ht]
$$\psfig{figure=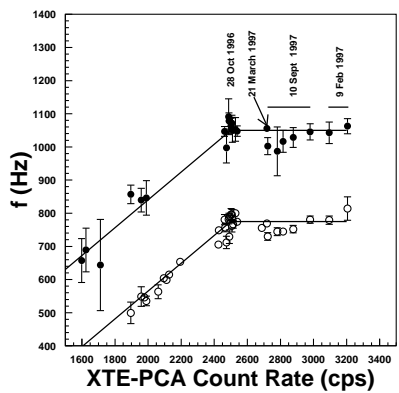,height=2.5in,angle=-90}$$
\caption{Evidence for a leveling off of the kHz QPO frequency with
count rate in 4U\,1820$-$30. (Zhang et al. 1998b) \label{fig:1820ISCO}}
\end{figure}

The possibility to detect evidence in kHz QPOs for the existence of the
innermost stable circular orbit (ISCO) predicted by general relativity,
which would constitute the first direct detection of a strong-field
general-relativistic effect, has fascinated since the beginning (Kaaret,
Ford and Chen 1997; Zhang, Strohmayer and Swank 1997). It has been
conjectured that when the inner edge of the accretion disk reaches
the ISCO, the QPO frequency might level off and remain constant while
\mdot\ continues rising. For this reason, the recent
measurement of an apparent leveling off of the increase of QPO
frequency with X-ray count rate in 4U\,1820$-$30 (Zhang et al. 1998b;
Fig.~\ref{fig:1820ISCO}) attracted considerable attention.

\begin{figure}[ht]
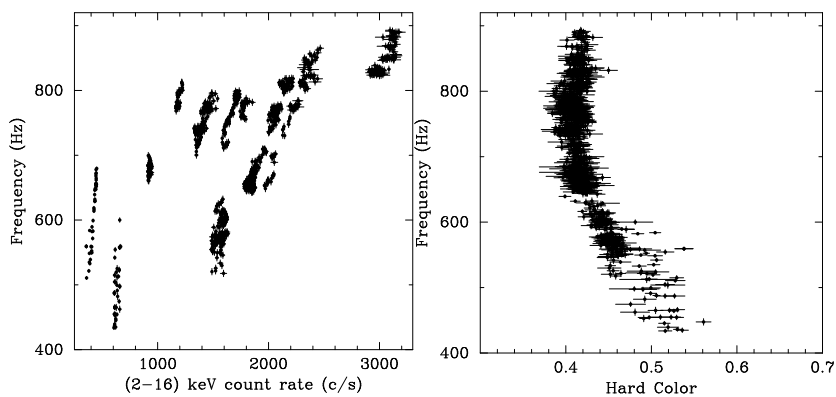

$$\psfig{figure=mendez_1608_apjlxxx_fig2_mess.ps,height=2in,angle=-90}
\psfig{figure=mendez_1608_apjlxxx_fig3_order.ps,height=2in,angle=-90}$$
\caption{In 4U\,1608$-$52 a QPO frequency vs. count rate plot shows no
clear correlation. When plotted vs. X-ray color it is clear there is
no saturation of QPO
frequency. (M\'endez et al. 1998d) \label{fig:1608freqvsint}\label{fig:1608freqvscol}}
\end{figure}

The difficulty in measurements of this kind is not in measuring the
frequency, but in measuring \mdot. We have known since the days of
EXOSAT that variations in X-ray count rate or even X-ray flux in the
band accessible to our instruments do not necessarily track \mdot\
(Hasinger and van der Klis 1989, van der Klis et al. 1990, Hasinger et
al. 1990), and RXTE observations have confirmed this beautifully. A
clear example of this is seen in 4U\,1608$-$52 (M\'endez et al. 1998d;
Fig.~\ref{fig:1608freqvsint} left). This figure, containing many more
measurements, looks much less nice than Fig.~\ref{fig:1820ISCO}. With
sparser coverage it could easily have happened to give a similar
impression of leveling off. When plotting frequency vs. X-ray color,
which in this case may be a better measure of \mdot\ (though not a
perfect one either), there is no sign of any saturation
(Fig.~\ref{fig:1608freqvscol}).

\begin{figure}[ht]
$$\psfig{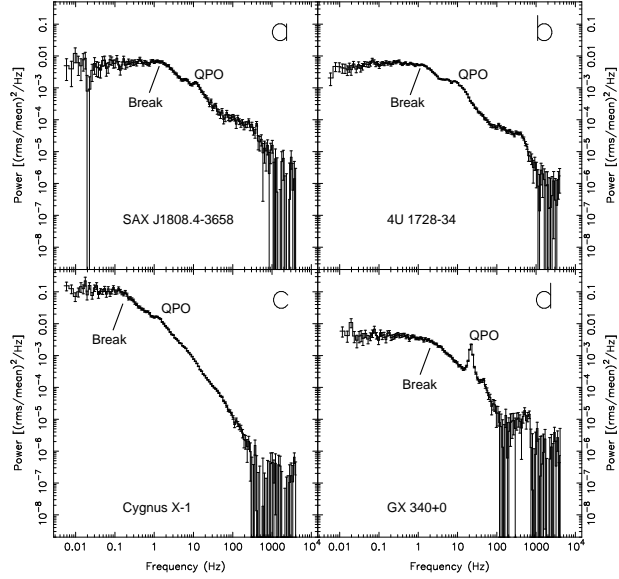}$$
\caption{Broad-band power spectra of, respectively, the millisecond pulsar, an atoll
source, a black-hole candidate and a Z source. (Wijnands and van der
Klis 1998d) \label{fig:wijnandsspectra}}
\end{figure}

\begin{figure}[ht]
$$\psfig{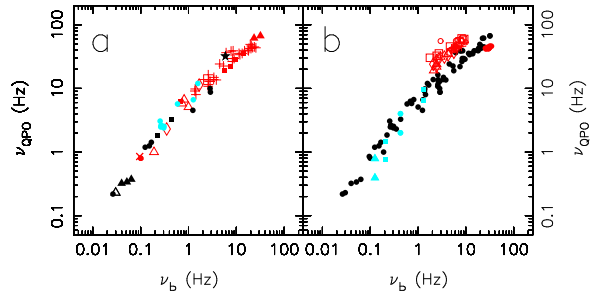}$$
\caption{Relation between noise break frequency and QPO frequency for
sources of the types shown in the previous figure. (Wijnands and van der
Klis 1998d)\label{fig:wijnandsbbn}}
\end{figure}

A number of intriguing correlations has been found between the kHz
QPOs and the phenomena at lower frequencies. Wijnands and van der Klis
(1998d) point out, that the low \mdot\ power-spectral similarity
between BHCs and low-luminosity low-magnetic field neutron stars found
with EXOSAT and Ginga (van der Klis 1994) also holds for the
millisecond pulsar SAX\,J1808.4-3658 (\S\,\ref{subsect:mspulsar}), and
may even extend to Z sources in {\it their} lowest \mdot\ states (the
left end of the so-called horizontal branch), where by the way these
sources are still quite luminous. The power spectra look very similar
(Fig.~\ref{fig:wijnandsspectra}), and with the better RXTE data it is now evident that
nearly always, in addition to the break at low frequency, a QPO-like
feature above the break is present. The correlation between the break
frequency and the QPO frequency is excellent
(Fig.~\ref{fig:wijnandsbbn}), and encompasses both neutron stars and
black-hole candidates. This suggests that both the band-limited noise
component with the 0.03--30\,Hz break frequency, and the 0.2--70\,Hz
QPO (with the possible exception of the Z sources, which are slightly
off the main relation) are found in both neutron stars
and black holes. This would exclude spin-orbit beat-frequency models
and any other models requiring a material surface, an event horizon, a
magnetic field, or their absence, for their explanation, essentially
implying these phenomena are generated in the accretion flow towards
{\it any} low-magnetic field compact object and are most likely disk
variability features. Ford and van der Klis (1998) studied the
relation between both of these features (the band-limited noise and
the low-frequency QPO-like feature) and the kHz QPOs in 4U\,1728$-$34,
and find generally good correlations (there is one deviant data set),
suggesting that kHz QPOs also fit into schemes of this kind in some
way. This of course leaves one with the question how the absence of
kHz QPOs of the type discussed in this article from black hole
candidates fits in with this (the highest-frequency QPOs seen in black
hole candidates have frequencies up a few 100\,Hz, but although
the amount of information on these QPOs is relatively limited, they
appear to have properties that are different from kHz QPOs in neutron
stars).

\begin{figure}[ht]
$$\psfig{figure=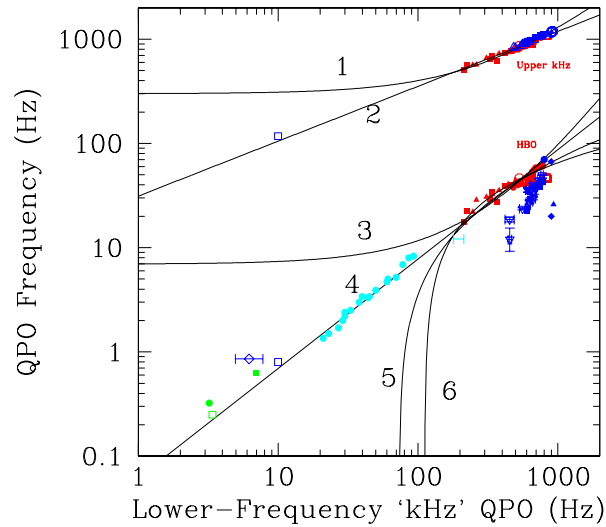,height=3.5in,angle=-90}$$
\caption{When plotting the frequencies of various QPO and broad noise
components seen in accreting neutron stars and black holes vs. each
other, a diagram emerges that suggests that similar components occur
in all these sources and span a very wide range in
frequency. Power-law relations describe the correlations reasonably
well. The curved lines illustrate various attempts to fit these
relations with spin-orbit beat-frequency models at the same spin
frequency for all sources. Such models always converge to a constant
frequency and therefore do not fit. (Psaltis, Belloni and van der Klis
1998) \label{fig:psaltisbellonivdklis}}
\end{figure}

A coincidence of QPO properties pointed out by Psaltis, Belloni and
van der Klis (1998) may provide an answer to this question. Power
spectra of many Z and atoll sources, of Cir\,X-1 and of a few low
luminosity neutron stars and BHCs show two QPO or broad noise
phenomena whose centroid frequencies, when plotted vs. each other,
seem to line up (Fig.~\ref{fig:psaltisbellonivdklis}). This would not
only identify the low-frequency (few 10\,Hz) QPOs in Z and atoll
sources with even lower-frequency QPOs in Cir\,X-1 (3--10\,Hz) and
BHCs (0.3--1\,Hz) as also suggested by the Wijnands and van der Klis
(1998d) results, but would {\it also} indicate that the {\it lower}
kHz peak in Z and atoll sources (the same argument does not hold for
the upper kHz peak), the broad 20--100\,Hz bumps found in Cir\,X-1
(Shirey et al. 1996, 1998, see also Tennant 1987), and even
lower-frequency (0.2--1\,Hz) bumps in some low luminosity neutron
stars and BHCs are due to the same physical phenomenon. While this
conjecture remains to be confirmed, the currently available data are
suggestive. The interpretation would be similar to that mentioned
above, particularly, it would imply that the lower kHz QPO is not
unique to neutron star systems either, and therefore presumably a
general feature of the accretion flow onto a compact object. Of
course, {\it orbital motion} in the disk is such an accretion flow
phenomenon and remains an attractive interpretation for some of the
observed frequencies. The phenomenology is quite complex; further
detailed work will show to what extent these various suggestions will
stand the test of additional data.

\section{Some final remarks}

A number of new theoretical ideas has been proposed for explaining
various aspects of the phenomenology described in the previous
sections. Examples are the papers by Alpar and Y\i lmaz (1997), Ghosh
(1998), Klu\'zniak (1998), Titarchuk et al. (1998). An interesting
discussion has also begun about the possibility, suggested by the
beat-frequency interpretation of the kHz QPOs, that all these neutron
stars spin at approximately the same rate (Bildsten 1998, Anderson et
al. 1998, Levin 1998).

For the purpose of these final remarks let us concentrate on one
aspect of the recent discussions and postulate that any succesful
model should be able to expain at least: (i) the apparent
beat-frequency relation between the twin kHz peaks and the burst
oscillations (Table~\ref{tab:beatevidence}), (ii) the relations
between kHz QPOs and lower-frequency QPOs in Z and atoll sources
described above, and (iii) the systematic variations in kHz QPO peak
separation with \mdot\ (Fig.~\ref{fig:sco1608}). Here I concentrate on
what may be shaping up to be two rivals: the spin-orbit beat-frequency
interpretation, and the general-relativistic apsidal motion/precession
interpretation. In versions of the beat-frequency model (BFM)
considered by Miller et al. (1998) and Psaltis et al. (1998), the kHz
QPOs are identified with the orbital frequency at the sonic radius and
its beat with the neutron star spin, and the lower-frequency HBO in Z
sources with the beat of the orbital motion at the {\it
magnetospheric} radius with the same neutron star spin. In
general-relativistic (GR) precession/apsidal motion models discussed
by Stella and Vietri (1998a,b) the kHz peaks are due to orbital motion
and GR apsidal motion of a slightly eccentric orbit, and the
low-frequency QPO to Lense-Thirring precession of this same
orbit. Both types of model may be able, with some difficulty, to
explain (ii) (Psaltis et al. 1998). While of course BFMs are built
around, and so have no difficulty to explain, (i), in these models the
question is: how to explain, in a spin-orbit beat model, the varying
peak separation (iii). The GR models may be able to explain
(iii) (Stella and Vietri 1998b), but seem to have no obvious way to
predict a beat-frequency relation (i). Surprisingly, ways out may
exist from both dilemma's. In the sonic-point beat-frequency model,
the beat interaction between spin and orbital frequencies takes place
at the sonic radius by an interaction between an orbiting clump and an
X-ray beam of the central, unseen, pulsar. However, the signal
associated with this interaction is produced only when the matter
released from this clump arrives at the neutron star surface. As
pointed out by Lamb (1998, priv. comm.), if the clump is in an orbit
that is gradually spiraling down, then the observed frequency will be
higher than the actual beat frequency at which beam and clump
interact, because the propagation time for the matter from the clump
to the surface will be gradually diminishing. This will put the lower
kHz peak closer to the upper one, and thus decrease the kHz peak
separation, more so when due to stronger radiation drag at higher
$L_x$ the spiralling-down is faster, as observed.

In the GR apsidal motion model, the upper kHz peak is at the orbital
frequency and the lower one at the apsidal motion frequency of a
slightly eccentric orbit. Interestingly, the difference between these
two frequencies is the frequency at which the orbiting clump goes
through periastron. So, perhaps in a model of this type one could
explain the fact that the difference frequency between the kHz peaks
is sometimes seen back during X-ray bursts by some kind of interaction
at periastron with, e.g., an expanding bursting layer. Whether this
would lead to a viable model remains to be seen. In particular it
might be a problem to account for the apparent constancy of the burst
oscillation frequency in each source.

At the conclusion of my 1997 Lipari review to which the current text
is intended to be an update I said that ``eventually, most LMXBs will
likely exhibit the new phenomenon'', and that this would help
enormously in constraining the models. Indeed, this is turning out to
be the case. A further synthesis among what is now a considerable and
fortunately fast-growing body of data on the properties of kHz QPOs,
low-frequency QPOs and X-ray spectra observed simultaneously is likely
to bring much progress in the year to come. Further long RXTE
observations of these sources will be indispensible in reaching a full
understanding of the systematics of the phenomenology.

\def\lw{Lewin, W.H.G.}
\def\vpj{Van Paradijs, J.}
\def\mk{Van der Klis, M.}
\def\aj{{AJ}}                   % Astronomical Journal
\def\araa{{ARA\&A\ }}             % Annual Review of Astron and Astrophys
\def\apj{{ApJ\ }}                 % Astrophysical Journal
\def\apjl{{ApJ\ }}                % Astrophysical Journal, Letters
\def\apjs{{ApJS\ }}               % Astrophysical Journal, Supplement
\def\ao{{Appl.~Opt.}}           % Applied Optics
\def\apss{{Ap\&SS}}             % Astrophysics and Space Science
\def\aap{{A\&A\ }}                % Astronomy and Astrophysics
\def\aapr{{A\&A~Rev.}}          % Astronomy and Astrophysics Reviews
\def\aaps{{A\&AS}}              % Astronomy and Astrophysics, Supplement
\def\azh{{AZh}}                 % Astronomicheskii Zhurnal
\def\baas{{BAAS}}               % Bulletin of the AAS
\def\jrasc{{JRASC}}             % Journal of the RAS of Canada
\def\memras{{MmRAS}}            % Memoirs of the RAS
\def\mnras{{MNRAS}}             % Monthly Notices of the RAS
\def\pra{{Phys.~Rev.~A}}        % Physical Review A: General Physics
\def\prb{{Phys.~Rev.~B}}        % Physical Review B: Solid State
\def\prc{{Phys.~Rev.~C}}        % Physical Review C
\def\prd{{Phys.~Rev.~D}}        % Physical Review D
\def\pre{{Phys.~Rev.~E}}        % Physical Review E
\def\prl{{Phys.~Rev.~Lett.}}    % Physical Review Letters
\def\pasp{{PASP}}               % Publications of the ASP
\def\pasj{{PASJ}}               % Publications of the ASJ
\def\qjras{{QJRAS}}             % Quarterly Journal of the RAS
\def\skytel{{S\&T}}             % Sky and Telescope
\def\solphys{{Sol.~Phys.}}      % Solar Physics
\def\sovast{{Soviet~Ast.}}      % Soviet Astronomy
\def\ssr{{Space~Sci.~Rev.}}     % Space Science Reviews
\def\zap{{ZAp}}                 % Zeitschrift fuer Astrophysik
\def\nat{{Nat\ }}              % Nature
\def\iaucirc{{IAU~Circ.}}       % IAU Cirulars
\def\aplett{{Astrophys.~Lett.}} % Astrophysics Letters
\def\apspr{{Astrophys.~Space~Phys.~Res.}}
                % Astrophysics Space Physics Research
\def\bain{{Bull.~Astron.~Inst.~Netherlands}} 
                % Bulletin Astronomical Institute of the Netherlands
\def\fcp{{Fund.~Cosmic~Phys.}}  % Fundamental Cosmic Physics
\def\gca{{Geochim.~Cosmochim.~Acta}}   % Geochimica Cosmochimica Acta
\def\grl{{Geophys.~Res.~Lett.}} % Geophysics Research Letters
\def\jcp{{J.~Chem.~Phys.}}      % Journal of Chemical Physics
\def\jgr{{J.~Geophys.~Res.}}    % Journal of Geophysics Research
\def\jqsrt{{J.~Quant.~Spec.~Radiat.~Transf.}}
                % Journal of Quantitiative Spectroscopy and Radiative Trasfer
\def\memsai{{Mem.~Soc.~Astron.~Italiana}}
                % Mem. Societa Astronomica Italiana
\def\nphysa{{Nucl.~Phys.~A}}   % Nuclear Physics A
\def\physrep{{Phys.~Rep.}}   % Physics Reports
\def\physscr{{Phys.~Scr}}   % Physica Scripta
\def\planss{{Planet.~Space~Sci.}}   % Planetary Space Science
\def\procspie{{Proc.~SPIE}}   % Proceedings of the SPIE

\end{document}